# Towards diluted magnetism in TaAs

Yu Liu,[1,§] Zhilin Li,[2,§] Liwei Guo,[2] Xiaolong Chen,[2,*] Ye Yuan,[1,4] Chi Xu,[1,4] René Hübner,[1] Shavkat Akhmadaliev,[1] Arkady V. Krasheninnikov,[1] Alpha T. N'Diaye,[3] Elke Arenholz,[3] Manfred Helm,[1,4] Shengqiang Zhou[1]

[1]Helmholtz-Zentrum Dresden-Rossendorf, Institute of Ion Beam Physics and Materials Research, Bautzner Landstraße 400, 01328 Dresden, Germany

[2]Research & Development Center for Functional Crystals, Beijing National Laboratory for Condensed Matter Physics, Institute of Physics, Chinese Academy of Sciences, Beijing 100190, China

[3]Advanced Light Source, Lawrence Berkeley National Laboratory, Berkeley, California 94720, USA

[4]Technische Universität Dresden, 01062 Dresden, Germany

[§]Both authors contributed equally to this work.

[*]chenx29@iphy.ac.cn

Magnetism in Weyl semimetals is desired to investigate the interaction between the magnetic moments and Weyl fermions, e.g. to explore anomalous quantum Hall phenomena. Here we demonstrate that proton irradiation is an effective tool to induce ferromagnetism in the Weyl semimetal TaAs. The intrinsic magnetism is observed with a transition temperature above room temperature. The magnetic moments from $d$ states are found to be localized around Ta atoms. Further, the first-principles calculations indicate that the $d$ states localized on the nearest-neighbor Ta atoms of As vacancy sites are responsible for the observed magnetic moments and the long-ranged magnetic order. The results show the feasibility of inducing ferromagnetism in Weyl semimetals so that they may facilitate the applications of this material in spintronics.

Over the last decade, numerous studies of the low-energy spectrum of graphene and topological insulators showed that electronic excitations in these systems can be described in terms of high-energy particles predicted by the quantum field theory, which provides a potential tabletop research platform for concepts borrowed from particle physics. For instance, the linear dispersion of electrons in graphene [1] and topological insulators [2, 3] allows to describe electrons as massless Dirac fermions. These quasiparticles were confirmed to exist in the three-dimensional space [4-6] soon after Dirac semimetals with fourfold degenerate Dirac nodes were predicted [7, 8].

Weyl fermions [9] are expected when a Dirac node is designed to be divided into two overlapping nodes with opposite chirality in semimetals, e.g. the transition-metal mono-pnictides TaAs, TaP, NbAs, and NbP [10, 11], which are referred to as Weyl semimetals (WSMs). So far, these WSMs have attracted extensive attention. Both the Weyl nodes and the Fermi arcs have been experimentally observed in all mono-pnictides: TaAs [12-15], NbAs [16], TaP [17] and NbP [18]. The spin texture of surface Fermi arcs has been further confirmed in TaAs [19]. The nearly full spin polarization has been observed on the Fermi arc surface [20]. Chiral-anomaly-induced negative magnetoresistance has been measured in NbP [21], TaAs [22], and TaP [23]. Mobility has been found as high as $5 \times 10^6$ cm$^2$ V$^{-1}$ s$^{-1}$ at 1.85 K in NbP [21] and $1.8 \times 10^5$ cm$^2$ V$^{-1}$ s$^{-1}$ at 10 K in TaAs [22]. Most recently, magnetic quantum-oscillations have been revealed in NbP [24]. These WSMs are diamagnetic to keep the Berry curvature intact [10, 25, 26]. As Weyl fermions can be seen as magnetic monopoles in the reciprocal space, an important question is how the spin coupling will affect WSMs, or whether their properties can be tuned by magnetic impurities and domains for potential applications. For example, ferromagnetism in WSMs may give rise to anomalous quantum Hall effect, which could promote the development of spintronic applications [27]. Therefore, the understanding of how magnetism can be induced in WSMs and its origin is of great significance. Sessi et al. [28] have investigated the magnetic perturbations on the TaAs surface after introducing Mn atoms. However, contrary to the surface of the material, it is much more difficult to realize magnetism in its bulk by post-synthesis methods, while the impurities may hinder or complicate the synthesis. At the same time, as Ta atoms in TaAs have spin-pairing $d^2$ electrons, a perturbation of the crystal field, e.g. by the presence of vacancies, may lead to spin polarization of these $d^2$ electrons.

In this work, we use proton irradiation to create vacancies and thus to modify the crystal field in the Weyl semimetal TaAs. Ferromagnetism emerges after proton irradiation, and as our first-principles calculations indicate, the local magnetic moments from the *d* states localized on the nearest-neighbor Ta atoms of As vacancies ($V_{As}$) are responsible for the observed long-range magnetic order.

**MATERIALS AND METHODS**

**Materials.** TaAs crystals were grown using the chemical vapor transport method. The pre-reacted TaAs polycrystalline precursor mixed with the transporting medium iodine was loaded into a quartz tube. After evacuation and sealing, the samples were kept at the growth temperature of around 1300 K for three weeks [29]. Single crystallites were grown, and some residual polycrystalline precursor could be detected. After basic cleaning, X-ray diffraction (XRD) and transmission electron microscopy (TEM) were used to confirm that the grown crystallites are of NbAs-type body-centered tetragonal structure. Fig. 1(a) shows an X-ray diffraction pattern of finely ground TaAs powder. The diffraction pattern is consistent with ICDD-PDF card No. 33-1388 for tetragonal TaAs. A more detailed XRD investigation on the TaAs samples was published in Ref. [29]. Fig. 1(b) shows a bright-field TEM image of a single TaAs crystallite. The diffraction contrast in the bright-field image is due to varying thickness and bending of the crystallite and most probably the presence of defects. In particular, stacking faults were observed in Ref. [30]. Fig. 1(c) shows the high-resolution TEM (HR-TEM) image from the area marked with a black square in Fig. 1(b). The corresponding fast Fourier transform (FFT), which is shown in Fig. 1(d), can be described with a [321] zone axis pattern of tetragonal TaAs. In Ref. [30], the authors have shown that despite the high density of defects in TaAs, the Weyl fermion pockets near the Fermi surface are well preserved. It should be mentioned, that no iodine doping or second phase formation are detected within the sensitivity limits. More details related to the sample preparation were reported previously [29].

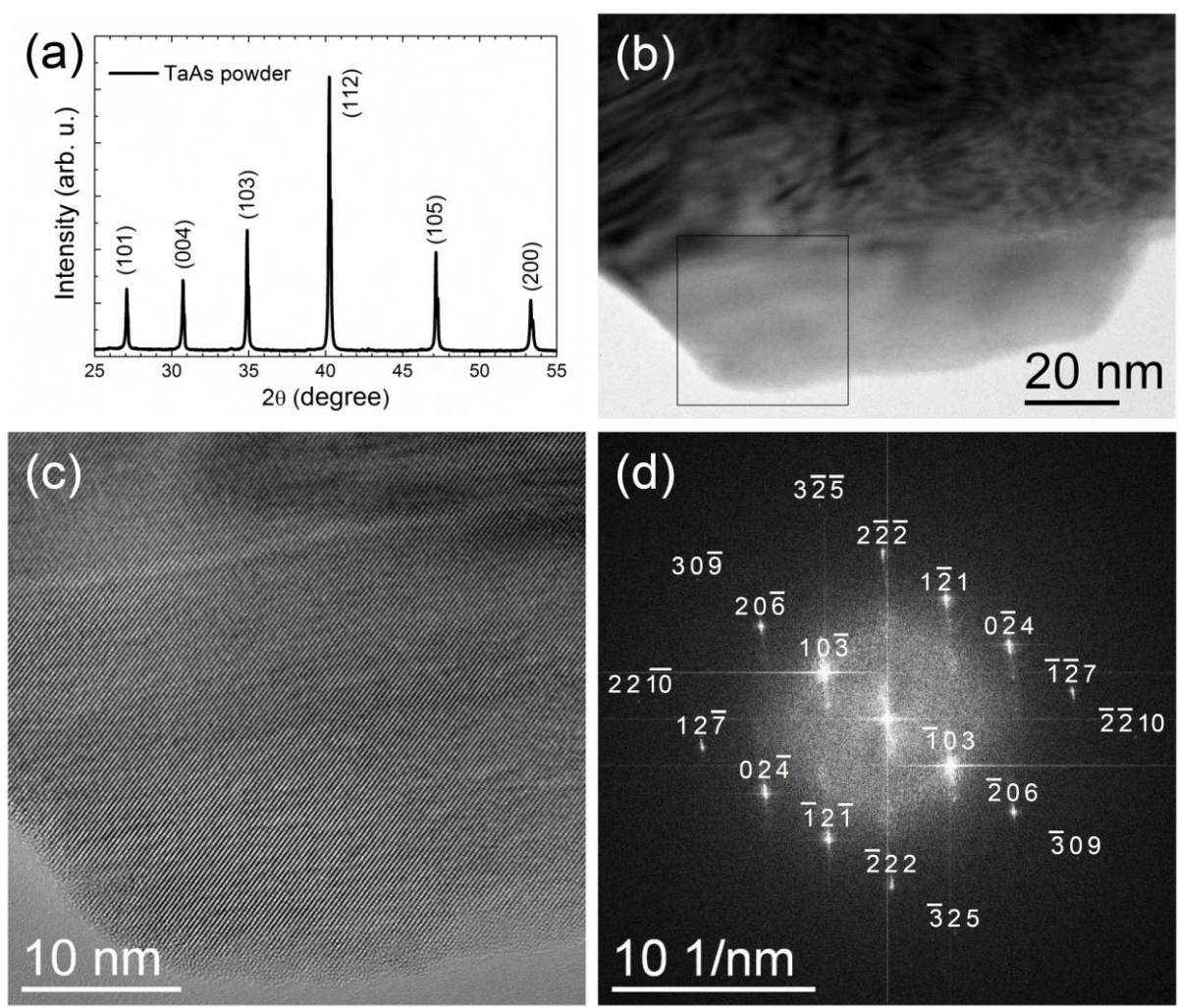

Figure 1. (a) XRD pattern of finely ground TaAs powder with indexing based on the NbAs-type structure (space group $I4_1md$), (b) bright-field TEM image of a TaAs crystallite, (c) HR-TEM image from the area marked with a black square in (b), (d) fast Fourier transform of the HR-TEM image in (c) with indexing based on a [321] zone axis pattern of tetragonal TaAs.

**Proton irradiation.** Proton irradiation was implemented using a 6 MV Tandem accelerator in the ion beam center at Helmholtz-Zentrum Dresden-Rossendorf. TaAs samples was irradiated by H ions with the incident energy of 12 MeV. The samples with the H fluences varying from $1 \times 10^9$ to $1 \times 10^{13}$ cm$^{-2}$ are labeled according to their fluences. Hydrogen and defect distributions predicted using SRIM (Stopping and Range of Ions in Matter) simulation are demonstrated in Fig. S1. As the thickness of all the samples is less than 0.3 mm, the distribution of defects is nearly uniform.

**Characterization.** The magnetization properties were measured by using the vibrating sample magnetometer attached a commercial superconducting quantum interference device (SQUID-VSM, Quantum Design) with a sensitivity limit of $10^{-7}$ emu. Both field-cooled (FC) and zero-field-cooled (ZFC) temperature dependent magnetization (M-T) measurements were performed from 5 to 395 K with the applied field from 200 to 2000 Oe. Magnetization versus applied magnetic field (M-H) at 5 and 300 K was measured in the field range of 5 kOe. X-ray absorption spectrum (XAS) and X-ray magnetic circular dichroism (XMCD) spectroscopy measurements were performed at BL 6.3.1 of the Advanced Light Source by using total electron yield (TEY) mode under a magnetic field of 0.5 T at 80 K for Ta $M_5$-edge and for As $L_3$-edge, respectively. The as-obtained results have been cross-checked by field reversal for each photon energy while the X-ray polarization remained constant.

**First-principles calculations.** To understand the magnetic properties of proton irradiated TaAs, we performed first-principles calculations on the basis of density-functional theory with spin polarization involved. All calculations were carried out by using the Perdew-Burke-Ernzerhof exchange-correlation functional based on the generalized gradient approximation [31] in the Cambridge Serial Total Energy Package [32]. The cutoff energy was set to 330 eV for the plane wave basis to represent the self-consistently treated valence electrons, as ultrasoft pseudopotentials depicted the core-valence interaction [33]. Self-consistent field calculations had the tolerance of $5.0 \times 10^{-7}$ eV/atom. The first Brillouin zone was sampled by dividing the grid spacing according to Monkhorst-Pack special k-point scheme of 0.04 Å$^{-1}$ for both wave functions and the density of states (DOS) and 0.015 Å$^{-1}$ for the band structure [34]. Supercell consisting of $2 \times 2 \times 1$, $3 \times 3 \times 1$, $4 \times 4 \times 1$ unit cells of TaAs containing one neutral As/Ta vacancies ($V_{As}$ or $V_{Ta}$) were built for calculations, corresponding to a defect concentration of 6.25%, 2.78%, 1.56%, respectively.

**RESULTS**

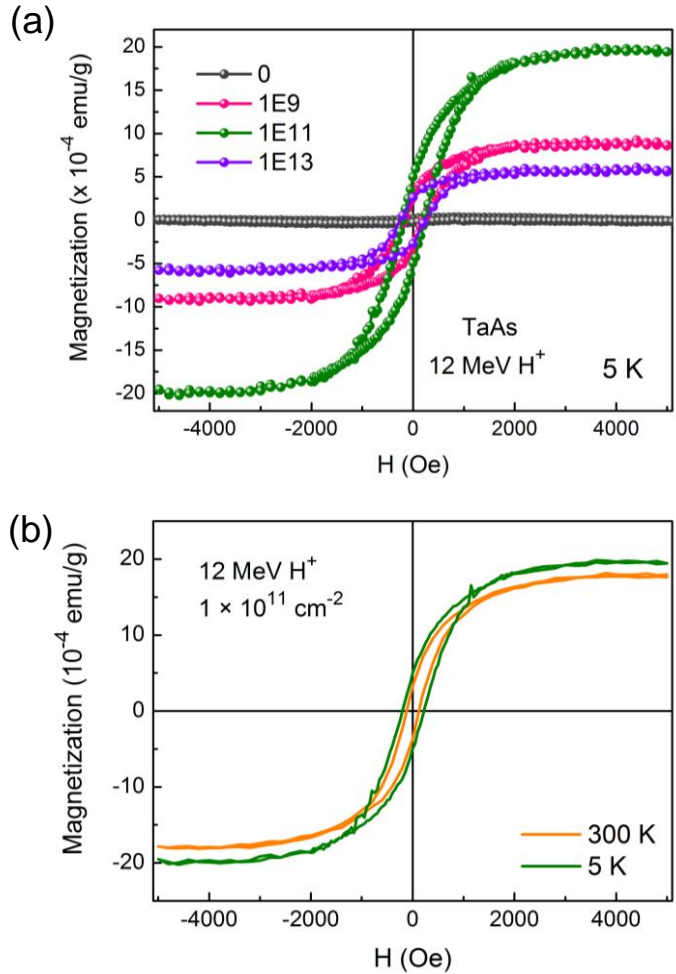

FIG. 2. The magnetization vs. field measurements. (a) The magnetization as a function of magnetic field without the diamagnetic background. It is performed within a 5 kOe magnetic field at 5 K for irradiated TaAs samples. The saturation magnetization reaches $2 \times 10^{-3}$ emu/g in the sample irradiated with a fluence of $1 \times 10^{11}$ cm$^{-2}$. The rise and fall of saturation

magnetization upon increasing fluence indicate that the magnetic moments likely originate from defects created during irradiation. (b) The magnetization vs. field without the diamagnetic contribution in the magnetic field range of 5 kOe for the sample with fluence of 1 × $10^{11}$ cm$^{-2}$ at T = 5 K and 300 K for comparison, which shows that the transition temperature is above room temperature.

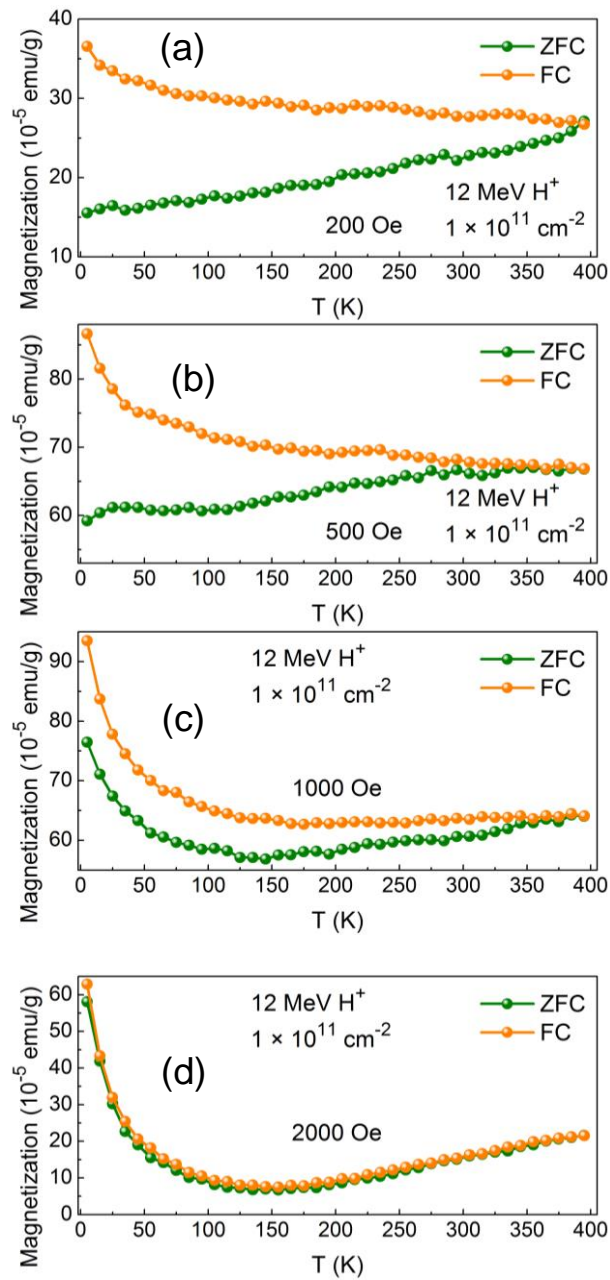

FIG. 3. ZFC/FC magnetization varying with temperature for the sample of 1 × $10^{11}$ cm$^{-2}$ from 5 to 395 K under a field of 200 Oe, 500 Oe, 1000 Oe or 2000 Oe, respectively. The ZFC/FC

curves show the FM feature at 200 and 500 Oe. Weak paramagnetism due to irradiation-induced damage is revealed. A minimum around 150 K appears, similar to that in the previous report [26].

The samples to be irradiated with protons are diamagnetic and similar to that reported in the previous work [26]. After proton irradiation, deviations from pure diamagnetism can be clearly observed in TaAs samples. After removing the diamagnetic background, the distinct hysteresis loops with ferromagnetic (FM) features are revealed in irradiated samples[Fig. 2(a)] and the saturation magnetization reaches $2 \times 10^{-3}$ emu/g in the sample irradiated with a fluence of $1 \times 10^{11}$ cm$^{-2}$ at 5 K. Even at 300 K, the saturation magnetization still remains about $1.7 \times 10^{-3}$ emu/g, as shown in Fig. 2(b). Therefore, its transition temperature is above room temperature. In contrast, the pristine sample shows no hysteresis as shown in Fig. 2(a). The absence of tiny parasitic or any secondary magnetic phase is confirmed by smooth and featureless temperature dependence of the magnetization curves (Fig. 3). Weak paramagnetism due to irradiation-induced damage is revealed as ZFC at 1000 and 2000 Oe and all FC magnetization curves rise along with temperature dropping. When the field reaches 2000 Oe, the ZFC/FC curves coincide with each other. A minimum around 150 K appears, similar to that in the previous report [26]. The rise and fall of saturation magnetization upon increasing fluence is observed, which is similar to that of magnetism induced by high-energy particle irradiation in various materials, such as ion implanted / neutron irradiated SiC and neutron irradiated Si [35-37]. These observations indicate that the magnetic moments likely originate from defects created during irradiation.

The XMCD spectroscopy can establish a direct connection between the ferromagnetism and the electrons/orbitals at the specific element [38, 39]. It is feasible even when the system has no magnetic elements. For example, the magnetic moments have been proven to originate from the $p_z$ orbitals at carbon atoms around divacancies in noble gas implanted SiC [40]. It has also been demonstrated that ferromagnetism in Si after neutron irradiation comes from 3s*3p* hybrid states of V6 [37]. Figure 4 shows the typical XAS and XMCD of the proton irradiated TaAs with a fluence of $1 \times 10^{13}$ cm$^{-2}$ at the Ta $M_5$-edge and at the As $L_3$-edge at 80 K, respectively. The XAS spectra are nearly the same before and after irradiation at the As $L_3$-edge. By a careful checking at the Ta $M_5$-edge around 1775 eV, one can notice a difference in the XAS spectrum, which we attribute to a change in the bonding of Ta *d* states. That would mean that Ta interacts strongly with irradiation induced defects. The XMCD signal at the As $L_3$-edge is vague, while it is detected as a peak at around 1770 eV at the Ta $M_5$-edge. This signal from Ta has peaks being different from that in SiC or Si [37, 40], but similar to that from 3d metals, e.g., Co or Mn in $Co_{1-x}Mn_x$ films or Mn in highly oriented pyrolytic graphite (HOPG) [41, 42]. Therefore we conclude that the *d* states localized on Ta lead to the formation of local moments in TaAs.

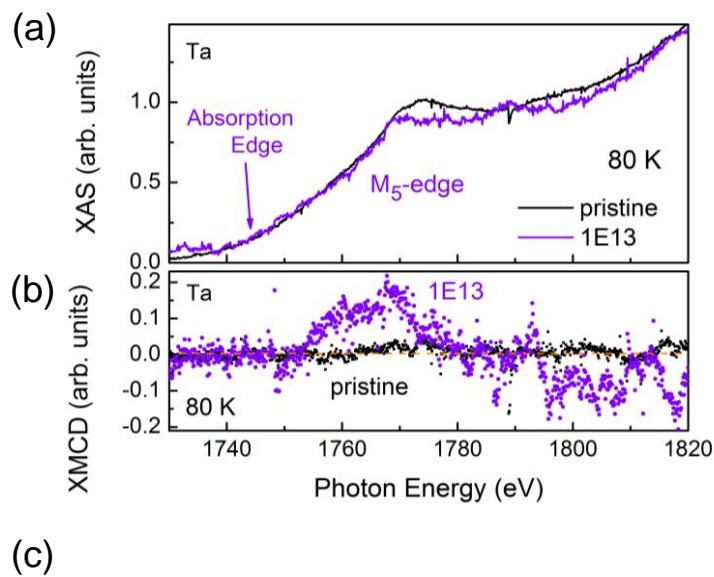

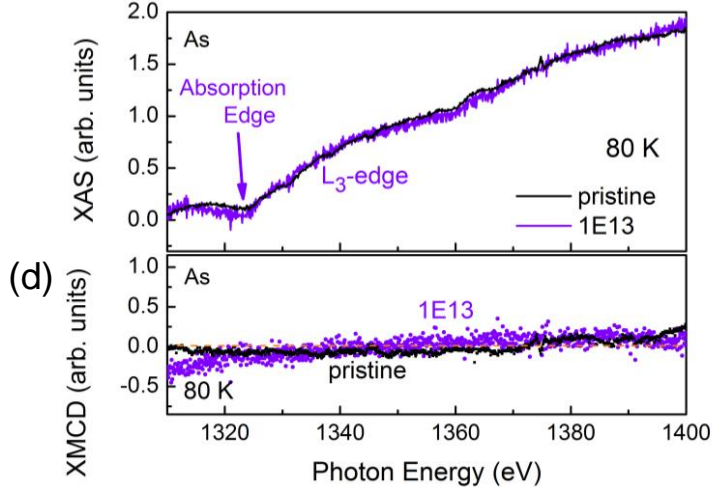

FIG. 4. The XAS and XMCD of pristine TaAs and proton irradiated TaAs with the fluence of $1 \times 10^{13}$ cm$^{-2}$ at the Ta M$_5$-edge and at the As L$_3$-edge at 80 K. (a) The XAS of the Ta M$_5$-edge, (b) The XMCD at the Ta M$_5$-edge, (c) The XAS of the As L$_3$-edge, (d) The XMCD at the As L$_3$-edge. XAS excites core level electrons into the conduction band, so it can probe the empty band structure. XMCD probes the element-specific magnetization using selection rules for excitation with circularly polarized light. For As, the absorption spectra are practically identical and there is no dichroism for As for either of the samples. For Ta, the XAS signatures of irradiated and pristine TaAs differ at around 1775 eV which is attributed to a change in the bonding of Ta $d$ states. The XMCD spectrum shows a peak at around 1770 eV after irradiation, which means the $d$ states of Ta contribute the magnetic moments.

**DISCUSSION**

Neither intentional doping nor detected magnetism is found in the as-grown samples, so that the magnetism must have been induced during the irradiating process. Proton irradiation will displace atoms in the lattice and create interstitials and vacancies like V$_{Ta}$ or V$_{As}$. As each proton creates many defects in the sample, the defects should be considered as the most likely cause for the observed magnetism. To understand the origin of the observed magnetism, we carried out first-principles calculations based on the spin-polarized density-functional theory. The calculated spin-resolved DOS of the 128-atom supercell [see Fig. 5] shows that each $V_{As}$ yields a magnetic moment of about 0.9 $\mu_B$. Vacancies do not lead to the appearance of a band gap, but give rise to the increase in DOS near the Fermi level, which is consistent with the previous results [30]. Therefore, the band splitting induced by spin polarization is not distinct between the majority- and minority-spin states. The partial DOS in Fig. 5 shows the spin polarization near the Fermi level mainly originates from $d$ states. Figure 6(a) shows the spin-resolved charge density isosurface of defect states associated with a $V_{As}$ in the 128-atom supercell. It indicates that the magnetic moments are mostly brought by the nearest-neighbor Ta atoms of $V_{As}$. The $d_{z^2}$ orbital feature is revealed at these Ta atoms. Our calculations for interstitials and Ta vacancies showed that the contribution of these defects to the observed magnetism is small.

The density of the local moments involved in ferromagnetism due to $V_{As}$ is estimated to be $7.7 \times 10^{17}$ $\mu_B$ cm$^{-3}$, so the $V_{As}$ concentration should be $8.5 \times 10^{17}$ cm$^{-3}$ at the fluence of $1 \times 10^{13}$ cm$^{-2}$. The SRIM calculations gave a considerably smaller concentration of irradiation induced As vacancies, than the one derived from the magnetic measurements. Specifically, depending on the parameters used (e.g. atom displacement value), it was found to be smaller by a factor of 2 or even order of magnitude. However, the irradiation-induced defects we did not consider, e.g., antisites, interstitial clusters, may have magnetic moments, so that with account for many uncontrollable approximations used in SRIM, one can say that the experimental results are in line with the theoretical estimates.

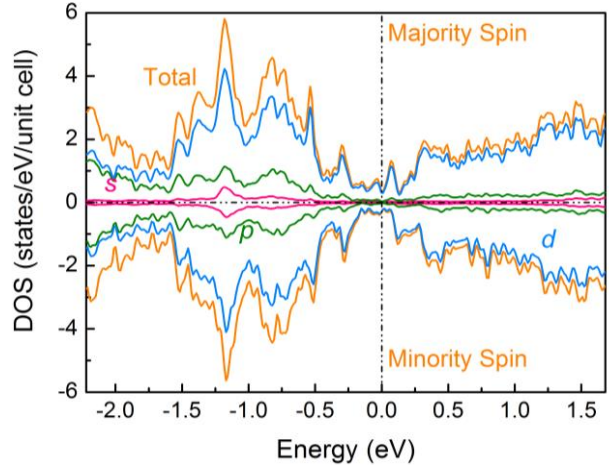

FIG. 5. Total and partial spin-resolved DOS of a 128-atom TaAs supercell with a $V_{As}$. The band splitting induced by spin polarization is not distinct. The spin polarization near the Fermi level mainly originates from $d$ states.

To study the long-range coupling (ferromagnetic or antiferromagnetic) between the $V_{As}$ local moments induced by the extended tails of the defect wave functions, the size of the supercell is doubled by putting two 128-atom supercells side by side. The energy difference between the antiferromagnetic (AFM) and FM phases is $E_{AFMab} - E_{FM} = 4J_{ab}S^2$ and $E_{AFMc} - E_{FM} = 4J_cS^2$ according to the nearest-neighbor Heisenberg model, where $E_{FM}$, $E_{AFMab}$ and $E_{AFMc}$ are the total energies of FM, AFM in the $a$-$b$ plane and AFM along the $c$ axis configurations, $J_{ab}$ and $J_c$ are the nearest-neighbor exchange interaction in and out of the $a$-$b$ plane, and S is the net spin of the defect states. A negative J means that antiferromagnetism is energetically favored and otherwise ferromagnetism favored. The magnetic configuration of $V_{As}$ with separation of 11.7 Å is ferromagnetism with $J_{ab}$ of 54.8 meV and $J_c$ of 0.7 meV. It is noted that $J_c$ is relatively small, so no spin ordering is preferred along the $c$ axis.

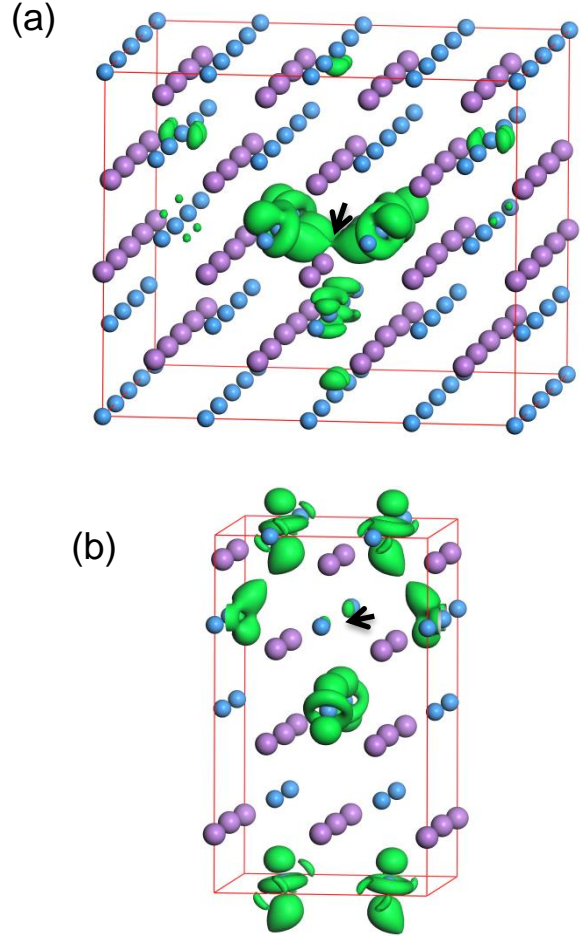

FIG. 6. Spin resolved isosurface charge density plot in dark green (a) of a 128-atom TaAs supercell with a $V_{As}$ (isovalue is 0.016 e/Å$^3$) and (b) of a 32-atom TaAs supercell with a $V_{Ta}$ (isovalue is 0.05 e/Å$^3$), demonstrating both the localized and the extended states due to the defects. Ta atoms are shown as smaller blue balls and As atoms are bigger and violet. It indicates that the magnetic moments are mostly brought by the nearest-neighbor Ta atoms of $V_{As}$, while the nearest-neighbor As atoms of $V_{Ta}$ donate no magnetic moments but most of moments are located around nearest-neighbor Ta atoms. Different from $V_{As}$ (The $d_{z^2}$ orbital feature), $V_{Ta}$ shows spin distribution with both $d_{z^2}$ and $d_{yz}$ orbital features. The arrows show the position of $V_{As}$.

Table 1. The spin-polarization energy ($E_{sp}$), magnetic moments ($M_s$), $J_{ab}$ and $J_c$ as a function of distance $d_V$ between the adjacent vacancies. $V_{As}$s are spin-polarized and have FM coupling

in the *a-b* plane regardless of the distance. $V_{Ta}$ have magnetic moments only when their concentration is high, about 3%, so that we assumed that their contribution to the observed magnetism is small.

| Supercell | Defect type | at% | $d_V$ (Å) | $E_{sp}$ (meV) | $M_s$ (µB) | $J_{ab}$ (meV) | $J_c$ (meV) |
|---|---|---|---|---|---|---|---|
| 2 × 2 × 1 | $V_{Ta}$ | 6.25 | 6.97 | 93.6 | 2.4 | 19.7 | 0.6 |
| 2 × 2 × 1 | $V_{As}$ | 6.25 | 6.97 | 4.7 | 0.3 | 101 | -17.2 |
| 3 × 3 × 1 | $V_{Ta}$ | 2.78 | 10.4 | 12.5 | 1.6 | -39.7 | -4.5 |
| 3 × 3 × 1 | $V_{As}$ | 2.78 | 10.4 | 93.4 | 1 | 33.3 | -3.7 |
| 4 × 4 × 1 | $V_{Ta}$ | 1.56 | 11.7 | 0 | | | |
| 4 × 4 × 1 | $V_{As}$ | 1.56 | 11.7 | 58 | 0.9 | 54.8 | 0.7 |

The variation of spin polarization and spin coupling is also investigated along with varying the distance between $V_{As}$s. As shown in Table 1, $V_{As}$s remain spin-polarized and still have FM coupling in the *a-b* plane regardless of the distance. However, the coupling along the *c* axis could be weak antiferromagnetism. We further considered the case of $V_{Ta}$ in TaAs. It is found that $V_{Ta}$ only obtains spin polarization when its concentration is high enough, about 3%. Figure 6(b) shows that the nearest-neighbor As atoms of $V_{Ta}$ donate no magnetic moments but most of moments are located around nearest-neighbor Ta atoms. The *d* orbital feature is also revealed at these Ta atoms. Different from $V_{As}$, $V_{Ta}$ shows spin distribution with both $d_{z^2}$ and $d_{yz}$ orbital features. In TaAs, Ta atoms have open *d* orbitals, so local moments will be more favorable to stay on these more localized orbitals contrast to *s* or *p* orbitals. Therefore, the ferromagnetism in proton irradiated TaAs can originate from spin-polarized electrons with *d* states located at Ta atoms close to $V_{As}$ sites according to the analysis mentioned above.

So far, we have studied the magnetic properties of the Weyl semimetal TaAs after proton irradiations. This method to induce magnetism could also be effective to other mono-pnictides, newly found type-II Weyl semimetal transition-metal dichalcogenides like $T_d$-$MoTe_2$ or $T_d$-$WTe_2$ [43, 44], and even proposed materials with unconventional quasiparticles [45]. The further investigation of transport properties is expected to reveal more details about the influence of internal magnetic field on the Weyl fermions, which will call for well-defined Weyl semimetal samples. As the Weyl nodes and the Fermi arcs in TaAs have been verified, the influence of magnetism in Weyl semimetals on the observation of electronic surface states using angle resolved photoemission spectroscopy can be studied. The change of Berry curvature because of magnetism can also be explored. Numerous challenging issues are waiting for answers, and their exploration is possible now because magnetism can successfully be introduced in Weyl semimetals. Moreover, the development of the top-down method for Weyl semimetals based on focused ion beam microfabrication may give birth to genuine topological devices [46-48].

**CONCLUSIONS**

In summary, we have investigated the magnetism of the Weyl semimetal TaAs induced by 12 MeV proton irradiation with different fluences. The magnetization can be as high as $2 \times 10^{-3}$ emu/g in the sample with the fluence of $1 \times 10^{11}$ cm$^{-2}$ at 5 K. Even at 300 K, the saturation magnetization remains almost the same as that at 5 K. Irradiation has almost no influence on the electronic states related to As atoms, while it changes those around Ta atoms. The magnetic moments from *d* states are found to be localized around Ta atoms based on XMCD. According to first-principles calculations, the $d_{z^2}$ states located on the nearest-neighbor Ta atoms of $V_{As}$ are responsible for the observed local moments which have long-range FM order within the *a-b* plane. The results show the feasibility to induce ferromagnetism without involving impurities in Weyl semimetals. These findings will deepen the understanding of TaAs properties and facilitate the development of Weyl semimetals for spintronic devices.

**ACKNOWLEDGEMENTS**

Y.L. would like to thank Dr. Anna Semisalova, Dr. Zhitao Zhang, Dr. Hannes Kühne, Fang Liu, Mao Wang, Johannes Klotz from HZDR, Prof. Dr. Fei Ding and Yan Chen from IFW Dresden for their assistance during the measurements, Prof. Dr. Carsten Timm from TU Dresden, Prof. Binghai Yan and Dr. Chandra Shekhar from MPI-CPfS, and Dr. Slawomir Prucnal from HZDR for fruitful discussion and Ning Liu from institute of physics, CAS for the assistance with the calculations. The work is financially supported by the Helmholtz Postdoc Programme (Initiative and Networking Fund, PD-146), the Ministry of Science and Technology of China (Grants No. 2011CB932700), the National Natural Science Foundation of China (Grants Nos. 51532010, 91422303, 51202286 and 51472265), and the Strategic Priority Research Program (B) of the Chinese Academy of Sciences (Grant No. XDB07020100). Support by the Ion Beam Center (IBC) at HZDR is gratefully acknowledged. The Advanced Light Source is supported by the Director, Office of Science, Office of Basic Energy Sciences, of the U.S. Department of Energy under Contract No. DE-AC02-05CH11231.

**REFERENCES**

[1]     K. S. Novoselov, A. K. Geim, S. V. Morozov, D. Jiang, M. I. Katsnelson, I. V. Grigorieva, S. V. Dubonos, and A. A. Firsov, Nature **438**, 197 (2005).

[2]     M. Z. Hasan and C. L. Kane, Rev. Mod. Phys. **82**, 3045 (2010).

[3]     X. L. Qi and S. C. Zhang, Rev. Mod. Phys. **83**, 1057 (2011).

[4]     Z. K. Liu, B. Zhou, Y. Zhang, Z. J. Wang, H. M. Weng, D. Prabhakaran, S. K. Mo, Z. X. Shen, Z. Fang, X. Dai, Z. Hussain, and Y. L. Chen, Science **343**, 864 (2014).

[5]     S. Y. Xu, C. Liu, S. K. Kushwaha, R. Sankar, J. W. Krizan, I. Belopolski, M. Neupane, G. Bian, N. Alidoust, T. R. Chang, H. T. Jeng, C. Y. Huang, W. F. Tsai, H. Lin, P. P. Shibayev, F. C. Chou, R. J. Cava, and M. Z. Hasan, Science **347**, 294 (2015).

[6]     Z. K. Liu, J. Jiang, B. Zhou, Z. J. Wang, Y. Zhang, H. M. Weng, D. Prabhakaran, S. K. Mo, H. Peng, P. Dudin, T. Kim, M. Hoesch, Z. Fang, X. Dai, Z. X. Shen, D. L. Feng, Z. Hussain, and Y. L. Chen, Nat. Mater. **13**, 677 (2014).

[7]     Z. J. Wang, Y. Sun, X. Q. Chen, C. Franchini, G. Xu, H. M. Weng, X. Dai, and Z. Fang, Phys. Rev. B **85**, 195320 (2012).

[8]     Z. J. Wang, H. M. Weng, Q. S. Wu, X. Dai, and Z. Fang, Phys. Rev. B **88**, 125427 (2013).

[9]     H. Weyl, Z. Phys. **56**, 330 (1929).

[10]    H. M. Weng, C. Fang, Z. Fang, B. A. Bernevig, and X. Dai, Phys. Rev. X **5**, 011029 (2015).

[11]    S. M. Huang, S. Y. Xu, I. Belopolski, C. C. Lee, G. Q. Chang, B. K. Wang, N. Alidoust, G. Bian, M. Neupane, C. L. Zhang, S. Jia, A. Bansil, H. Lin, and M. Z. Hasan, Nat. Commun. **6**, 7373 (2015).

[12]    B. Q. Lv, H. M. Weng, B. B. Fu, X. P. Wang, H. Miao, J. Ma, P. Richard, X. C. Huang, L. X. Zhao, G. F. Chen, Z. Fang, X. Dai, T. Qian, and H. Ding, Phys. Rev. X **5**, 031013 (2015).

[13]    S. Y. Xu *et al.*, Science **349**, 613 (2015).

[14]    B. Q. Lv, N. Xu, H. M. Weng, J. Z. Ma, P. Richard, X. C. Huang, L. X. Zhao, G. F. Chen, C. E. Matt, F. Bisti, V. N. Strocov, J. Mesot, Z. Fang, X. Dai, T. Qian, M. Shi, and H. Ding, Nat. Phys. **11**, 724 (2015).

[15] L. X. Yang, Z. K. Liu, Y. Sun, H. Peng, H. F. Yang, T. Zhang, B. Zhou, Y. Zhang, Y. F. Guo, M. Rahn, D. Prabhakaran, Z. Hussain, S. K. Mo, C. Felser, B. Yan, and Y. L. Chen, Nat. Phys. **11**, 728 (2015).

[16] S. Y. Xu *et al.*, Nat. Phys. **11**, 748 (2015).

[17] N. Xu *et al.*, Nat. Commun. **7**, 11006 (2016).

[18] Z. K. Liu, L. X. Yang, Y. Sun, T. Zhang, H. Peng, H. F. Yang, C. Chen, Y. Zhang, Y. F. Guo, D. Prabhakaran, M. Schmidt, Z. Hussain, S. K. Mo, C. Felser, B. Yan, and Y. L. Chen, Nat. Mater. **15**, 27 (2016).

[19] B. Q. Lv, S. Muff, T. Qian, Z. D. Song, S. M. Nie, N. Xu, P. Richard, C. E. Matt, N. C. Plumb, L. X. Zhao, G. F. Chen, Z. Fang, X. Dai, J. H. Dil, J. Mesot, M. Shi, H. M. Weng, and H. Ding, Phys. Rev. Lett. **115**, 217601 (2015).

[20] S. Y. Xu *et al.*, Phys. Rev. Lett. **116**, 096801 (2016).

[21] C. Shekhar, A. K. Nayak, Y. Sun, M. Schmidt, M. Nicklas, I. Leermakers, U. Zeitler, Y. Skourski, J. Wosnitza, Z. K. Liu, Y. L. Chen, W. Schnelle, H. Borrmann, Y. Grin, C. Felser, and B. H. Yan, Nat. Phys. **11**, 645 (2015).

[22] X. C. Huang, L. X. Zhao, Y. J. Long, P. P. Wang, D. Chen, Z. H. Yang, H. Liang, M. Q. Xue, H. M. Weng, Z. Fang, X. Dai, and G. F. Chen, Phys. Rev. X **5**, 031023 (2015).

[23] F. Arnold, C. Shekhar, S. C. Wu, Y. Sun, R. D. dos Reis, N. Kumar, M. Naumann, M. O. Ajeesh, M. Schmidt, A. G. Grushin, J. H. Bardarson, M. Baenitz, D. Sokolov, H. Borrmann, M. Nicklas, C. Felser, E. Hassinger, and B. H. Yan, Nat. Commun. **7**, 11615 (2016).

[24] J. Klotz, S. C. Wu, C. Shekhar, Y. Sun, M. Schmidt, M. Nicklas, M. Baenitz, M. Uhlarz, J. Wosnitza, C. Felser, and B. H. Yan, Phys. Rev. B **93**, 121105 (2016).

[25] B. Roy and J. D. Sau, Phys. Rev. B **92**, 125141 (2015).

[26] Y. Liu, S. Prucnal, S. Q. Zhou, Z. L. Li, L. W. Guo, X. L. Chen, Y. Yuan, F. Liu, and M. Helm, J. Magn. Magn. Mater. **408**, 73 (2016).

[27]    C.-Z. Chang *et al.*, Science **340**, 167 (2013).

[28]    P. Sessi, Y. Sun, T. Bathon, F. Glott, Z. Li, H. Chen, L. Guo, X. Chen, M. Schmidt, C. Felser, B. Yan, and M. Bode, Phys. Rev. B **95**, 035114 (2017).

[29]    Z. L. Li, H. X. Chen, S. F. Jin, D. Gan, W. J. Wang, L. W. Guo, and X. L. Chen, Cryst. Growth Des. **16**, 1172 (2016).

[30]    T. Besara, D. A. Rhodes, K. W. Chen, S. Das, Q. R. Zhang, J. Sun, B. Zeng, Y. Xin, L. Balicas, R. E. Baumbach, E. Manousakis, D. J. Singh, and T. Siegrist, Phys. Rev. B **93**, 245152 (2016).

[31]    J. P. Perdew, K. Burke, and M. Ernzerhof, Phys. Rev. Lett. **77**, 3865 (1996).

[32]    S. J. Clark, M. D. Segall, C. J. Pickard, P. J. Hasnip, M. J. Probert, K. Refson, and M. C. Payne, Z. Kristallogr. **220**, 567 (2005).

[33]    D. Vanderbilt, Phys. Rev. B **41**, 7892 (1990).

[34]    H. J. Monkhorst and J. D. Pack, Phys. Rev. B **13**, 5188 (1976).

[35]    L. Li, S. Prucnal, S. D. Yao, K. Potzger, W. Anwand, A. Wagner, and S. Q. Zhou, Appl. Phys. Lett. **98**, 222508 (2011).

[36]    Y. T. Wang, Y. Liu, E. Wendler, R. Hubner, W. Anwand, G. Wang, X. L. Chen, W. Tong, Z. R. Yang, F. Munnik, G. Bukalis, X. L. Chen, S. Gemming, M. Helm, and S. Q. Zhou, Phys. Rev. B **92**, 174409 (2015).

[37]    Y. Liu, R. Pan, X. Zhang, J. Han, Q. Yuan, Y. Tian, Y. Yuan, F. Liu, Y. Wang, A. T. N'Diaye, E. Arenholz, X. Chen, Y. Sun, B. Song, and S. Zhou, Phys. Rev. B **94**, 195204 (2016).

[38]    G. Schütz, W. Wagner, W. Wilhelm, P. Kienle, R. Zeller, R. Frahm, and G. Materlik, Phys. Rev. Lett. **58**, 737 (1987).

[39]    J. Stöhr, H. A. Padmore, S. Anders, T. Stammler, and M. R. Scheinfein, Surf. Rev. Lett. **05**, 1297 (1998).

[40]  Y. T. Wang, Y. Liu, G. Wang, W. Anwand, C. A. Jenkins, E. Arenholz, F. Munnik, O. D. Gordan, G. Salvan, D. R. T. Zahn, X. L. Chen, S. Gemming, M. Helm, and S. Q. Zhou, Sci. Rep. **5**, 8999 (2015).

[41]  S. Guchhait, H. Ohldag, E. Arenholz, D. A. Ferrer, A. Mehta, and S. K. Banerjee, Phys. Rev. B **88**, 174425 (2013).

[42]  R. J. Snow, H. Bhatkar, A. T. N'Diaye, E. Arenholz, and Y. U. Idzerda, J. Magn. Magn. Mater. **419**, 490 (2016).

[43]  A. A. Soluyanov, D. Gresch, Z. Wang, Q. Wu, M. Troyer, X. Dai, and B. A. Bernevig, Nature **527**, 495 (2015).

[44]  Y. Sun, S.-C. Wu, M. N. Ali, C. Felser, and B. Yan, Phys. Rev. B **92**, 161107 (2015).

[45]  B. Bradlyn, J. Cano, Z. Wang, M. G. Vergniory, C. Felser, R. J. Cava, and B. A. Bernevig, Science **353**, aaf5037 (2016).

[46]  P. J. Moll, B. Zeng, L. Balicas, S. Galeski, F. F. Balakirev, E. D. Bauer, and F. Ronning, Nat. Commun. **6**, 6663 (2015).

[47]  P. J. W. Moll, P. Kushwaha, N. Nandi, B. Schmidt, and A. P. Mackenzie, Science **351**, 1061 (2016).

[48]  P. J. W. Moll, N. L. Nair, T. Helm, A. C. Potter, I. Kimchi, A. Vishwanath, and J. G. Analytis, Nature **535**, 266 (2016).